\documentclass[pra,english,preprintnumbers,amsmath,amssymb,nofootinbib,twocolumn,superscriptaddress]{revtex4-1}
\usepackage[utf8]{inputenc}
\usepackage[T1]{fontenc}
\usepackage{amsmath}
\usepackage{amssymb}
\usepackage{graphicx}
\usepackage{physics}
\usepackage{tabularx}
\usepackage{bm}
\usepackage{pifont}
\usepackage[colorlinks]{hyperref}

\usepackage{color}
\definecolor{orange}{rgb}{1,0.5,0}

\definecolor{bblue}{rgb}{0.2,0.8,1.0}

\newcommand{\beq}{\begin{equation}}
\newcommand{\eeq}{\end{equation}}
\newcommand{\bea}{\begin{eqnarray}}
\newcommand{\eea}{\end{eqnarray}}

\begin{document}

\title{Leading- and next-to-leading order semiclassical approximation to the first seven virial coefficients of 
spin-1/2 fermions across spatial dimensions}

\author{Y. Hou}
\affiliation{Department of Physics and Astronomy, University of North Carolina, Chapel Hill, North Carolina 27599, USA}

\author{A. J. Czejdo}
\affiliation{Department of Physics and Astronomy, University of North Carolina, Chapel Hill, North Carolina 27599, USA}

\author{J. DeChant}
\affiliation{Department of Physics and Astronomy, University of North Carolina, Chapel Hill, North Carolina 27599, USA}

\author{C. R. Shill}
\affiliation{Department of Physics and Astronomy, University of North Carolina, Chapel Hill, North Carolina 27599, USA}


\author{J. E. Drut}
\affiliation{Department of Physics and Astronomy, University of North Carolina, Chapel Hill, North Carolina 27599, USA}

\date{\today}

\begin{abstract}
Following up on recent calculations, we investigate the leading- and next-to-leading order semiclassical approximation
to the virial coefficients of a two-species fermion system with a contact interaction. Using the analytic result 
for the second-order virial coefficient as a renormalization condition, we derive expressions for up to the 
seventh-order virial coefficient $\Delta b_7$. Our results at leading order, though approximate, furnish 
simple analytic formulas that relate $\Delta b_n$ to $\Delta b_2$ for arbitrary dimension, providing a glimpse
into the behavior of the virial expansion across dimensions and coupling strengths.
As an application, we calculate the pressure and Tan's contact of the 2D attractive Fermi gas and examine
the radius of convergence of the virial expansion as a function of the coupling strength.
\end{abstract}

\maketitle 

\section{Introduction} 

In a recent paper~\cite{ShillDrut}, two of us presented results for virial coefficients in a semiclassical lattice 
approximation (SCLA), at leading order (LO), for spin-$1/2$ fermions with a contact two-body interaction. 
We found that, in spite of the crudeness of the approximation, the results for $\Delta b_3$ and $\Delta b_4$ 
were surprisingly good when written in terms of the exact $\Delta b_2$, which amounted to using the latter as 
a renormalized coupling. Specifically, quantitative or at least qualitative agreement was found between the
LO-SCLA and diagrammatic and Monte Carlo results for 1D and 2D Fermi gases with attractive 
interactions. 

In this work, we explore the LO-SCLA further by carrying out the evaluation of virial coefficients 
up to $b_7$, and furthermore extending our previous analysis to next-to-leading order (NLO). Our (approximate) 
analytic answers, obtained partially by algebra automation, provide insight 
into the behavior of the virial expansion as a function of the spatial dimension of the problem. As an application, we 
compare with Monte Carlo results for the density equation of state of an attractive 2D Fermi gas. 
As an additional example, we calculate the many-body contribution to Tan's contact for the same system.

\section{Hamiltonian and virial expansion}

We assume a non-relativistic kinetic energy and a two-body contact interaction, such that the 
Hamiltonian for two flavors $\uparrow,\downarrow$ is $\hat H = \hat T + \hat V$,
where
\bea
\label{Eq:T}
\hat T \!=\! {\int{d^d x\,\hat{\psi}^{\dagger}_{s}({\bf x})\left(-\frac{\hbar^2\nabla^2}{2m}\right)\hat{\psi}_{s}({\bf x})}},
\eea
and
\bea
\label{Eq:V}
\hat V \!=\! - g_d\! \int{d^d x\,\hat{n}_{\uparrow}({\bf x})\hat{n}_{\downarrow}({\bf x})},
\eea
where the field operators $\hat{\psi}_{s}, \hat{\psi}^{\dagger}_{s}$ are fermionic fields for particles of spin $s=\uparrow,\downarrow$ (summed over $s$ above), and $\hat{n}_{s}({\bf x})$ are the coordinate-space densities. In the remainder of this work, we will 
take $\hbar = k_\text{B} = m = 1$ and discretize spacetime using the spatial lattice spacing $\ell$ to set the scale for all quantities.
In particular, we will define the lattice kinetic energy exactly as above by using a momentum-space representation with periodic boundary conditions 
(rather than a local three-point formula for the second derivative in coordinate space), and the lattice potential energy will take the form
\bea
\label{Eq:Vlattice}
\hat V \!=\! - g_d\! \sum_{\bf x} \hat{n}_{\uparrow}({\bf x})\hat{n}_{\downarrow}({\bf x}),
\eea
where now all of the operators and constants on the right-hand side represent dimensionless lattice quantities, 
and we have omitted an overall prefactor $\ell^{-2}$ that gives $\hat V$ its physical units.

One way to characterize the thermodynamics of this system is through the virial expansion~\cite{VirialReview}, which is an expansion around the dilute limit
$z\to 0$, where $z=e^{\beta \mu}$ is the fugacity, i.e. it is a low-fugacity expansion. The corresponding coefficients
accompanying the powers of $z$ in the expansion of the grand-canonical potential $\Omega$ are the virial coeffiecients;
specifically,
\beq
-\beta \Omega = \ln {\mathcal Z} = Q_1 \sum_{n=1}^{\infty} b_n z^n,
\eeq
where 
\beq
\mathcal Z = \tr \left[e^{-\beta (\hat H - \mu \hat N)}\right] = \sum_{N=0}^{\infty} z^N Q_N,
\eeq 
is the grand-canonical partition function and $Q_N$ is the $N$-body partition function. By definition, $b_1 = 1$ and the higher-order 
coefficients require solving the corresponding few-body problems:
\bea
Q_1 b_2 &=& Q_2 - \frac{Q_1^2}{2!},\\
Q_1 b_3 &=& Q_3 - b_2 Q_1^2  - \frac{Q_1^3}{3!},\\
Q_1 b_4 &=& Q_4 -  \left(b_3 + \frac{b_2^2}{2}\right) Q_1^2 -b_2\frac{Q_1^3}{2!}  - \frac{Q_1^4}{4!},
\eea
and so forth (see Appendix).
For completeness and future reference, we note here the values of the noninteracting virial coefficients for 
nonrelativistic fermions in $d$ spatial dimensions: $b^{(0)}_n = (-1)^{n+1} n^{-(d+2)/2}$.

For our system, in arbitrary spatial dimensions,
\beq
Q_1 = 2\sum_{\bf p} e^{- \beta \frac{{\bf p}^2}{2m}},
\eeq
which in the continuum limit becomes $Q_1 = 2{V}/{\lambda_T^d}$, where $V$ is the $d$-dimensional spatial volume and 
$\lambda_T = \sqrt{2 \pi \beta}$ is the de Broglie thermal wavelength.
Since $Q_1 \propto V$, the above expressions for $b_n$ display precisely how the volume dependence should cancel out to
yield volume-independent coefficients.
In particular, the highest power of $Q_1$ does not involve the interaction and therefore always disappears in the interaction-induced change $\Delta b_n$:
\bea
Q_1 \Delta b_2 &=& \Delta Q_2, \\
Q_1 \Delta b_3 &=& \Delta Q_3 - Q_1^2 \Delta b_2,\\
Q_1 \Delta b_4 &=& \Delta Q_4 -  \Delta\left(b_3 + \frac{b_2^2}{2}\right) Q_1^2 -\frac{\Delta b_2}{2} Q_1^3,
\eea
and so forth. In the Appendix we show the corresponding expressions up to $b_7$, but the pattern is repeated: $\Delta b_n$ involves a contribution from $\Delta Q_n$
and several contributions involving the previous virial coefficients $b_m$, $m<n$ and powers of $Q_1$; the latter always cancel against specific terms within $\Delta Q_n$
to yield a volume-independent $b_n$. Those cancellations are a challenging feature for stochastic approaches, 
but they become a useful check for our calculations.

In terms of the partition functions $Q_{MN}$ of $M$ particles of one type and $N$ of the other type, we have
\bea
\Delta Q_2 &=& \Delta Q_{11}, \\
\Delta Q_3 &=& 2\Delta Q_{21}, \\
\Delta Q_4 &=& 2\Delta Q_{31} + \Delta Q_{22}, \\
\Delta Q_5 &=& 2\Delta Q_{41} + 2\Delta Q_{32}, \\
\Delta Q_6 &=& 2\Delta Q_{51} + 2\Delta Q_{42}+ \Delta Q_{33}, \\
\Delta Q_7 &=& 2\Delta Q_{61} + 2\Delta Q_{52}+ 2\Delta Q_{43}.
\eea
We thus see that the number of non-trivial contributions to each virial coefficient is actually small. The challenge is in determining each of these terms and for 
that purpose we implement the semiclassical approximation advertised above, which we describe in detail next.

\section{The semiclassical approximation at leading order}

\subsection{Basic formalism}
In a wide range of many-body methods, the grand-canonical partition function $\mathcal Z$ is expressed as a path integral over an auxiliary Hubbard-Stratonovich field.
Here we use a different route, but with the same first step: we introduce a Trotter-Suzuki (TS) factorization of the Boltzmann weight. 
At the lowest non-trivial order in such a factorization, 
\beq
e^{-\beta (\hat T + \hat V)} = e^{-\beta \hat T}e^{-\beta \hat V},
\eeq
where the higher orders involve exponentials of nested commutators of $\hat T$ with $\hat V$. Thus, the LO in this
expansion consists in setting $[\hat T , \hat V] = 0$, which becomes exact in the limit where either $\hat T$
or $\hat V$ can be ignored (i.e. respectively the strong- and weak-coupling limits). Orders
beyond LO can be reached using a factorization based on the Trotter identity
\beq
e^{-\beta (\hat T + \hat V)} = \lim_{n \to \infty} \left( e^{-\beta \hat T/n}e^{-\beta \hat V/n } \right)^n.
\eeq
Indeed, the leading
order can simply be viewed as the most coarse possible TS factorization, i.e. with time step $\tau = \beta$. 
Higher orders $n > 1$ will be defined by using progressively finer discretizations $\tau = \beta/n$.
We leave such explorations to future work.

\subsection{A simple example}
As the simplest example of the LO-SCLA, we calculate $Q_{11}$:
\bea
Q_{11} &=& \sum_{{\bf p}_1 {\bf p}_2} \langle {\bf p}_1 {\bf p}_2 | e^{-\beta \hat T}e^{-\beta \hat V} | {\bf p}_1 {\bf p}_2 \rangle \\
&&\!\!\!\!\!\!\!\!\!\!\!\! = \sum_{{\bf p}_1 {\bf p}_2} e^{-\beta (p_1^2 + p_2^2)/2m} \langle {\bf p}_1 {\bf p}_2 |e^{-\beta \hat V} | {\bf p}_1 {\bf p}_2 \rangle.
\eea
The kinetic energy operator piece is thus trivially evaluated. The central step is to insert a coordinate-space completeness relation 
to evaluate the potential energy piece, which we do using the following identity:
\bea
e^{-\beta \hat V} | {\bf x}_1 {\bf x}_2 \rangle &=& \prod_{\bf z} (1 + C\hat{n}_{\uparrow}({\bf z})\hat{n}_{\downarrow}({\bf z}))| {\bf x}_1 {\bf x}_2 \rangle  \\
&&\!\!\!\!\!\!\!\!\!\!\!\! = | {\bf x}_1 {\bf x}_2 \rangle + C \sum_{\bf z}\delta_{{\bf x}_1, {\bf z}} \delta_{{\bf x}_2, {\bf z}}| {\bf x}_1 {\bf x}_2 \rangle \nonumber \\
&&\!\!\!\!\!\!\!\!\!\!\!\! = \left[1 + C \delta_{{\bf x}_1, {\bf x}_2} \right]| {\bf x}_1 {\bf x}_2 \rangle \nonumber,
\eea
where $C = (e^{\beta g_d} - 1)\ell^d$ and we used the fermionic relation $\hat{n}^2_s = \hat{n}_s$. The $C$-independent term yields the noninteracting 
result, such that we may write
\beq
\!\!\!\!\Delta Q_{11} =  C\!\!\!\!\!\!\! \sum_{{\bf p}_1 {\bf p}_2,{\bf x}_1 {\bf x}_2} \!\!\!\!\!\!\!\! e^{-\beta (p_1^2 + p_2^2)/2m} \delta_{{\bf x}_1, {\bf x}_2} |\langle {\bf x}_1 {\bf x}_2 | {\bf p}_1 {\bf p}_2 \rangle |^2,
\eeq
which simplifies dramatically in this particular case when using a plane wave basis, because $|\langle {\bf x}_1 {\bf x}_2 | {\bf p}_1 {\bf p}_2 \rangle |^2 = 1/V^2$.
We then find
\beq
\Delta Q_{11} = C \frac{Q_{10}^2}{V},
\eeq
where
\beq
Q_{10} = \sum_{{\bf p}_1} e^{-\beta p_1^2/2m}.
\eeq
Thus, $\Delta b_2 = C {Q_{10}^2}/{(V Q_1)} = C {Q_{1}}/{(4V)}$, where we used $Q_1 = 2 Q_{10}$.
Following essentially the same steps, it is not difficult to see that
$\Delta b_3 = - C Q_1(2 \beta) / V$, where $Q_1(2 \beta)$ is $Q_1$ evaluated at $\beta \to 2\beta$.

\subsection{A more difficult example}

To display the complexity of the calculation in a less trivial case, we show $Q_{22}$ as another example.
Using the notation ${\bar {\bf P}} = ({\bf p}_1, {\bf p}_2, {\bf p}_3, {\bf p}_4)$, where $1,2$ refer to spin-up particles
and $3,4$ to spin-down particles, we have
\beq
Q_{22} = \sum_{{\bar {\bf P}} } e^{-\beta {\bar {\bf P}}^2/2m} \langle {\bar {\bf P}}  |e^{-\beta \hat V} | {\bar {\bf P}}  \rangle.
\eeq
As before, we must insert a complete set of coordinate eigenstates to evaluate the remaining matrix element.
To that end, we note that, using the notation ${\bar {\bf X}} = ({\bf x}_1, {\bf x}_2, {\bf x}_3, {\bf x}_4)$,
\bea
e^{-\beta \hat V} | \bar {\bf X}\rangle &=& [1+ C f_1(\bar {\bf X})  + C^2 f_2(\bar {\bf X})] | \bar {\bf X} \rangle,
\eea
where
\beq
f_1(\bar {\bf X}) = \delta_{{\bf x}_1, {\bf x}_3} \!+\! \delta_{{\bf x}_1, {\bf x}_4} \!+\!\delta_{{\bf x}_2, {\bf x}_3} \!+\! \delta_{{\bf x}_2, {\bf x}_4},
\eeq
and
\beq
f_2(\bar {\bf X}) = 2[\delta_{{\bf x}_1, {\bf x}_3}\delta_{{\bf x}_2, {\bf x}_4} + \delta_{{\bf x}_1, {\bf x}_4}\delta_{{\bf x}_2, {\bf x}_3}].
\eeq
Thus,
\beq
\Delta Q_{22} = \sum_{{\bar {\bf P}},{\bar {\bf X}} } e^{-\beta {\bar {\bf P}}^2/2m} |\langle {\bar {\bf X}} | {\bar {\bf P}}\rangle|^2
(C f_1(\bar {\bf X}) + C^2 f_2(\bar {\bf X})).
\eeq
Note that $\langle {\bar {\bf X}} | {\bar {\bf P}}\rangle$ factorizes across spins where, in this case,
each of the factors involved takes the form
\beq
|\langle {\bf y}_1 {\bf y}_2 | {\bf q}_1 {\bf q}_2 \rangle|^2 = \frac{1}{V^2}(1 - \cos(({\bf y}_1 -{\bf y}_2)({\bf q}_1 -{\bf q}_2))).
\eeq

Based on the above examples, it is easy to glean that the general form of the change $\Delta Q_{M,N}$ in the partition
function for $M$ spin-up particles and $N$ spin-down particles, with a contact interaction, is given by
\beq
\Delta Q_{MN} = \sum_{{\bar {\bf P}},{\bar {\bf X}} } e^{-\beta {\bar {\bf P}}^2/2m} |\langle {\bar {\bf X}} | {\bar {\bf P}}\rangle|^2
(C f_a(\bar {\bf X}) + C^2 f_b(\bar {\bf X}) + \dots),
\eeq
where ${\bar {\bf P}},{\bar {\bf X}}$ represent all momenta and positions of the $M+N$ particles, and the
functions $f_a$, $f_b$, $\dots$, which encode the matrix element of $e^{-\beta \hat V}$, depend on the specific 
case being considered. In particular, the case of $\Delta Q_{M1}$ is particularly simple and reduces to
\beq
\Delta Q_{M1} = M C\frac{Q_{10}}{V}\sum_{{\bar {\bf P}},{\bar {\bf X}} } e^{-\beta {\bar {\bf P}}^2/2m} 
|\langle {\bar {\bf X}} | {\bar {\bf P}}\rangle|^2,
\eeq
where ${\bar {\bf P}},{\bar {\bf X}}$ go over the momenta and positions of the $M$ identical particles.

The wavefunction $\langle {\bar {\bf X}} | {\bar {\bf P}}\rangle$ is a product of two Slater determinants which,
if using a plane-wave single-particle basis, leads to simple Gaussian integrals over the momenta $\bar {\bf P}$. 
The only challenge is in carrying out the sum over $\bar {\bf X}$ before the sum over $\bar {\bf P}$, as the
Slater determinants will naively lead to a large number of terms; to that end, it is crucial to use the interaction
matrix elements to simplify the determinants before carrying out any momentum sums or integrals.
In all cases, the integrals involved will be multidimensional Gaussian integrals.

\subsection{Generating $\mathcal O (C)$ results for all $n$}

As a check for our calculations and a useful result in itself, we show here how to calculate the 
contributions at $\mathcal O (C)$ in the LO-SCLA for all $n$. We begin 
with the grand-canonical partition function at LO, generalized to arbitrary chemical potentials
$\mu_\uparrow$ and $\mu_\downarrow$ (sum over $s = \uparrow, \downarrow$ implied below):
\bea
\mathcal Z &=& \tr \left[e^{-\beta (\hat T - \mu_s \hat N_s)} e^{-\beta \hat V}\right] \nonumber \\
&=& \tr \left[e^{-\beta (\hat T - \mu_s \hat N_s)} \prod_{\bf x} (1 + C\hat{n}_{\uparrow}({\bf x})\hat{n}_{\downarrow}({\bf x}))\right] \nonumber \\
&=& \mathcal Z_0^{}\left [ 1 + C \sum_{\bf x} \langle \hat{n}_{\uparrow}({\bf x})\hat{n}_{\downarrow}({\bf x}) \rangle^{}_0 + \mathcal O(C^2) \right],
\eea 
where $\mathcal Z_0^{} = \tr \left[e^{-\beta (\hat T - \mu_s \hat N_s)}\right]$ is the noninteracting partition function and 
$\langle \cdot \rangle^{}_0$ denotes a noninteracting thermal expectation value. This is, of course, nothing other than
leading-order perturbation theory. In the general, asymmetric case we have
\beq
\langle \hat{n}_{\uparrow}\hat{n}_{\downarrow} \rangle^{}_0 = 
\frac{1}{V^2}
\sum_{{\bf p},{\bf q}}  \frac{z_\uparrow e^{-\beta p^2/2m}}{1 + z_\uparrow e^{-\beta p^2/2m}}
 \frac{z_\downarrow e^{-\beta q^2/2m}}{1 + z_\downarrow e^{-\beta q^2/2m}}.
\eeq
By differentiation of the above expression with respect to $z_\uparrow = e^{\beta \mu_\uparrow}$ and 
$z_\downarrow = e^{\beta \mu_\downarrow}$, it is straightforward to derive the $\mathcal O (C)$ change 
in $\Delta Q_{M,N}$ for arbitrary $M,N$.

\section{Results}

\subsection{Virial coefficients on the lattice}

Using the above formalism, we obtained expressions for the virial coefficients on the lattice as shown in Tables 
\ref{Tab:Db2Db7C1SCLALattice}, \ref{Tab:Db2Db7C2SCLALattice}, and \ref{Tab:Db2Db7C3SCLALattice}, 
in the LO-SCLA. The results shown in those tables, up to $\Delta b_5$, were obtained on paper;
the remaining answers were obtained using an automated computer algebra code of our own design. 

As mentioned in the Introduction, the explicit calculation of $\Delta b_n$ by way of $\Delta Q_{MN}$ involves
delicate cancellations of various volume-dependent contributions which scale as $V$, $V^2$, $\dots, V^{n-1}$,
which yields a volume-independent result for $\Delta b_n$. As a check for our automated calculations, 
we have verified that those cancellations take place exactly as expected.

The functions $Q$, $F$, and $G$ appearing in \ref{Tab:Db2Db7C1SCLALattice}, \ref{Tab:Db2Db7C2SCLALattice}, and \ref{Tab:Db2Db7C3SCLALattice} are defined by
\beq
Q(x \beta) = \frac{2}{V} \sum_{\bf p} e^{- x\beta {\bf p}^2 / 2m},
\eeq
\beq
G({ M}) = \frac{1}{V^3} \sum_{{\bf p}_1 {\bf p}_2 {\bf p}_3} \exp(-\frac{\beta}{2 m} {\bf P}^T {\bf M} {\bf P}),
\eeq
\beq
F({ N}) = \frac{1}{V^4}\sum_{{\bf p}_1 {\bf p}_2 {\bf p}_3 {\bf p}_4} \exp(-\frac{\beta}{2 m} {\bf P}^T {\bf N} {\bf P}),
\eeq
where \( {\bf P}^T = ({\bf p}_1, {\bf p}_2, {\bf p}_3, \cdots) \) is a vector collecting all the $d$-dimensional momentum 
variables appearing in the sums, $\bf M$ and $\bf N$ are block matrices of 
the form
\beq
{\bf X} = 
    \begin{pmatrix}
      X_x & 0 & 0 \\
      0 & X_y & 0 \\
      0 & 0 & X_z \\
    \end{pmatrix},
\eeq
for $d=3$, where the explicit form of the block entries $X_i$ =  \( { M} \) and \( { N} \), respectively of size 
$3\times3$ and $4\times 4$, are shown in the Appendix. In the cases studied here,
the matrices $X_i$ will be the same for all cartesian components.

In the continuum limit, where the above sums turn into integrals, we have
\beq
Q(x \beta) \to \frac{2}{\lambda_T^d}\frac{1}{\sqrt{x}},
\eeq
\beq
G({ M}) \to \frac{1}{\lambda_T^{3d}} \left(\frac{1}{\det M}\right)^{d/2},
\eeq
and
\beq
F({ N}) \to \frac{1}{\lambda_T^{4d}} \left(\frac{1}{\det N }\right)^{d/2}.
\eeq
Using these formulas, we present our continuum results in the next section.

Note that, as a feature of the LO-SCLA, the expressions for $\Delta b_2$ and $\Delta b_3$ stop at $\mathcal O(C)$;
the results for $\Delta b_4$ and $\Delta b_5$ terminate at $\mathcal O(C^2)$; and finally
$\Delta b_6$ and $\Delta b_7$ go only up to $\mathcal O(C^3)$. We emphasize that those are full results
within the LO-SCLA rather than approximations in powers of $C$.

{\renewcommand{\arraystretch}{3}
\begin{table}
  \normalsize
    \centering
    \caption{\label{Tab:Db2Db7C1SCLALattice} \( \mathcal{O}(C) \) Terms for \( \Delta b_2 \) to \( \Delta b_7 \) }
    \begin{tabular}{c|c}
      \hline\hline
      \( \Delta b_n\) & \( \mathcal{O}(C) \) \\
      \hline\hline
      \( \Delta b_2 \) & \( \frac{Q(\beta)}{4}\) \\
      \hline
      \( \Delta b_3 \) & \( -\frac{Q(2\beta)}{2}\) \\
      \hline
      \( \Delta b_4 \) &
          \(\frac{Q(3\beta)}{2} + \frac{Q^2(2\beta)}{4 Q(\beta)} \) \\
      \hline
      \( \Delta b_5 \) &
          \(-\frac{Q(4\beta)}{2} - \frac{Q(2\beta) Q(3\beta)}{2 Q(\beta)}\) \\
      \hline
      \( \Delta b_6 \) &
          \(\frac{Q(5\beta)}{2}
              + \frac{Q(2\beta) Q(4\beta)}{2 Q(\beta)} 
              + \frac{Q(3\beta)^2}{4 Q(\beta)} \) \\
      \hline
      \( \Delta b_7 \) &
         \( - \frac{Q(6\beta)}{2}
            - \frac{Q(3\beta) Q(4\beta) + Q(2\beta) Q(5\beta)}{2 Q(\beta)}\)\\
      \hline\hline
    \end{tabular}
\end{table}
\begin{table}
  \normalsize
    \centering
    \caption{\label{Tab:Db2Db7C2SCLALattice} \( \mathcal{O}(C^{2}) \) Terms for \( \Delta b_2 \) to \( \Delta b_7 \) }
    \begin{tabular}{c|c}
      \hline\hline
      \( \Delta b_n\) &  \( \mathcal{O}(C^2) \) \\
      \hline\hline
      \( \Delta b_4 \) &
          \(\frac{G(M_4)}{2Q(\beta)} - \frac{Q(\beta) Q(2\beta)}{8} \) \\
      \hline
      \( \Delta b_5 \) &
          \(-2 \frac{G(M_5)}{Q(\beta)} + \frac{Q(\beta)Q(3\beta)}{4}
              + \frac{Q^2(2\beta)}{4}\) \\
      \hline
      \( \Delta b_6 \) &
          \(\begin{array}{c}
                \frac{2G(M_{6a}) + 2G(M_{6b}) + G(M_{6c})}{Q(\beta)}\\
              - \frac{3 Q(4\beta) Q(\beta)}{8}
              - \frac{Q^3(2\beta)}{8 Q(\beta)}
              - \frac{3 Q(3\beta) Q(2\beta)}{4} 
            \end{array} \) \\
      \hline
      \( \Delta b_7 \) &
         \(\begin{array}{c}
            - 2 \frac{3G(M^{(2)}_{7a}) + G(M^{(2)}_{7b}) + G(M^{(2)}_{7c})}{Q(\beta)}\\
            + \frac{Q^2(3\beta) + {Q(5\beta) Q(\beta)}}{2}\\
            + \frac{Q^2(2\beta)Q(3\beta)}{2 Q(\beta)}
            + Q(4\beta)Q(2\beta)
         \end{array}\) \\
      \hline\hline
    \end{tabular}
\end{table}

\begin{table}
  \normalsize
    \centering
    \caption{\label{Tab:Db2Db7C3SCLALattice} \( \mathcal{O}(C^3) \) Terms for \( \Delta b_2 \) to \( \Delta b_7 \) }
    \begin{tabular}{c|c}
      \hline\hline
      \( \Delta b_n\) & \( \mathcal{O}(C^3) \) \\
      \hline\hline
      \( \Delta b_6 \) &
          \(\begin{array}{c}
              - G(M_5)
              + \frac{2F(N_1)}{3 Q(\beta)}\\
              + \frac{Q^2(2\beta)Q(\beta)}{16}
              + \frac{Q(3\beta)Q^2(\beta)}{24} 
          \end{array}\)  \\
      \hline
      \( \Delta b_7 \) &
        \(\begin{array}{c}
              \frac{G(M^{(3)}_{7a}) Q(2\beta)}{Q(\beta)}
            - 4 \frac{F(N_2)}{Q(\beta)}\\
            + 3 G(M^{(3)}_{7b})
            + 2 G(M^{(3)}_{7c})\\
            - \frac{Q(4\beta) Q^2(\beta)
                    + 3 Q(3\beta) Q(2\beta) Q(\beta)
                    + Q^3(2\beta)}{8} 
        \end{array}\) \\
      
      \hline\hline
    \end{tabular}
\end{table}
}

\subsection{Virial coefficients in the continuum limit}
In Ref.~\cite{ShillDrut}, it was shown that the LO-SCLA gives
\bea
\label{Eq:Db3Db4SCLA}
\Delta b_3 &=& -2^{1-d/2}\Delta b_2, \\
\Delta b_4 &=&  2(3^{-d/2} + 2^{-d-1}) \Delta b_2 \nonumber \\ 
&& + 2^{1-d/2} \left(2^{-d/2-1} -1\right)(\Delta b_2)^2,
\eea
for a fermionic two-species system with a contact interaction, in $d$ spatial dimensions.
[Note that we have corrected the coefficient of $(\Delta b_2)^2$ relative to Ref.~\cite{ShillDrut}.]

The main result of this work is the extension of the above formulas to up to seventh order 
in the virial expansion and up to NLO in the SCLA. We collect the LO results in Table~\ref{Tab:Db3Db7Results}
and provide the NLO answers as Supplemental Material~\cite{SuppMatt}.
In the LO-SCLA, $\Delta b_2 \propto C$, and so $\Delta b_2$ tracks the order of $C$ appearing in 
each virial coefficient. At the same level in the approximation, $\Delta b_3$ only displays contributions up to
$\mathcal O(C)$, while $\Delta b_4$ and $\Delta b_5$ stop at $\mathcal O(C^2)$; 
$\Delta b_6$ and $\Delta b_7$ contain terms up to $\mathcal O(C^3)$.

\begin{table*}
  \normalsize
    \centering
    \caption{\label{Tab:Db3Db7Results} Full results in the LO-SCLA for \( \Delta b_3 \) through \( \Delta b_7 \) in 
    powers of $\Delta b_2$ in the continuum limit.}
    \begin{tabular}{c|c|c|c}
      \hline\hline
      \( \Delta b_n\) & \( \mathcal{O}( \Delta b_2) \)  & \( \mathcal{O}( \Delta b_2)^2 \)  & \( \mathcal{O}( \Delta b_2)^3 \)  \\
      \hline\hline
      \( \Delta b_3 \) & $-2^{1 - d/2}$  & --  & -- \\
      \hline
      \( \Delta b_4 \) & $ 2(3^{-d/2} + 2^{-d-1})$  &  $2^{1-d/2}(2^{-d/2-1} - 1)$  & -- \\
      \hline
      \( \Delta b_5 \) & $-2 (2^{-d} + 6^{-d/2})$  & $ 4 \left( 2^{-d} + 3^{-d/2} - 7^{-d/2} \right) $ & -- \\
      \hline
      \( \Delta b_6 \) & $2^{1-3d/2} + 3^{-d} + 2 \cdot 5^{-d/2}$ & 
      $
      \begin{array}{c}
        -2^{1-3 d / 2} - 3 \cdot 2^{1-d} (1 - 3^{-d/2}) \\
        -2^{2-d/2} ( 3^{1-d/2} -  5^{-d/2})
      \end{array}
      $  
                      & $
                        \begin{array}{c}
                          2^{2-d} -8 \cdot 7^{-d/2}\\
                          +8 \cdot 3^{-d/2-1} ( 1 + 2^{-d})
                        \end{array}
                        $ \\
      \hline
      \( \Delta b_7 \) & $ -2^{1-d/2} \left( 6^{-d/2} + 5^{-d/2} + 3^{-d/2} \right) $ & 
      $
      \begin{array}{c}
        2^{4 - 3 d/2} + 8 \cdot 5^{-d/2} \\
        + 8 \cdot  3^{-d/2} (3^{-d/2} + 2^{- d}) \\
        - 4 ( 2^{-d} 5^{-d/2} - 13^{-d/2} - 3 \cdot 17^{-d/2} )
     \end{array}
     $
                      & $
                        \begin{array}{c}
-2^{3- 3d/2} - 2^{3-d} ( 1 - 3^{1-d/2}) \\
-8 \cdot 2^{-d/2} (3^{1-d/2} - 7^{-d/2}) \\
-16 \cdot 10^{-d/2} (2^{-d/2} - 1)
                          \end{array}$ \\
      \hline\hline
    \end{tabular}
\end{table*}

\begin{figure}[t]
  \begin{center}
  \includegraphics[scale=0.56]{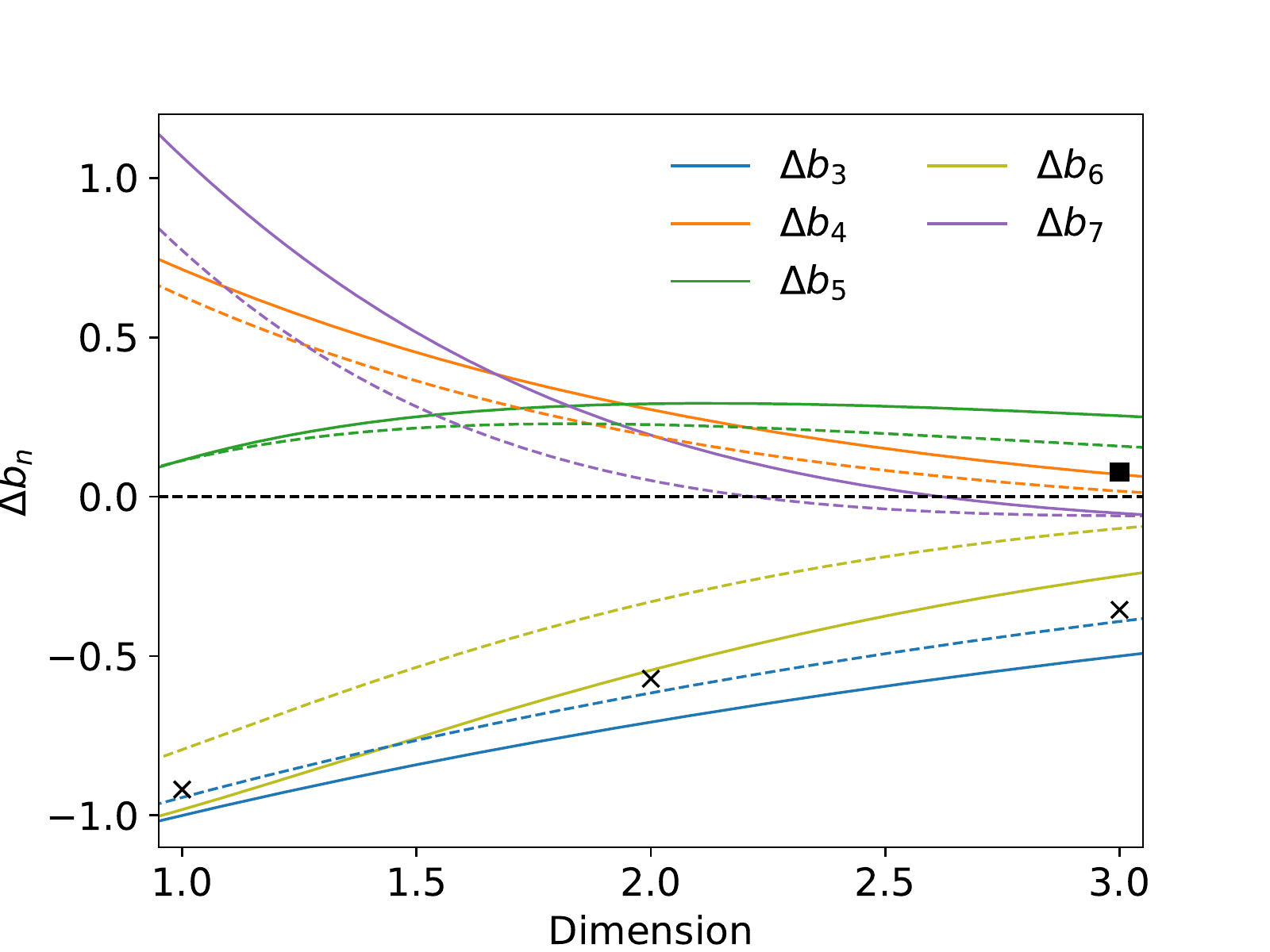}
  \end{center}
  \caption{Interaction-induced change in virial coefficients $\Delta b_3$ -- $\Delta b_7$
  at $\Delta b_2 = 1/\sqrt{2}$ (which corresponds to the unitary limit in 3D), as a function of the spatial 
  dimension $d$, in the LO- and NLO-SCLA (solid and dashed lines, respectively).
   The crosses represent $\Delta b_3$ as follows: Monte Carlo results in 1D from Ref.~\cite{ShillDrut}, 
   diagrammatic results in 2D from Ref.~\cite{Ngampruetikorn}, and exact results in 3D~\cite{Leyronas}.
   The square shows the $\Delta b_4$ result of Ref.~\cite{YanBlume}.
   }
  \label{Fig:DimensionComparison}
\end{figure}

The results in Table~\ref{Tab:Db3Db7Results} correspond to taking the continuum limit of the lattice expressions and using 
$\Delta b_2$ as the renormalized coupling to replace the bare lattice coupling $C$.
Using those results, we show in Fig.~\ref{Fig:DimensionComparison} a plot of $\Delta b_3, \dots, \Delta b_7$ as functions
of the spatial dimension $d$ at $\Delta b_2 = 1/\sqrt{2}$, which is the value corresponding to the 3D unitary Fermi gas. 
We compare those answers with the $\Delta b_3$ results of Ref.~\cite{ShillDrut} in 1D,
the diagrammatic results of Ref.~\cite{Ngampruetikorn} in 2D (see also Ref.~\cite{DrummondVirial2D}), and the known results in 3D (Ref.~\cite{Leyronas}
calculated the exact answer, while the work of Ref.~\cite{LiuHuDrummond} calculated it numerically, and Ref.~\cite{DBK} semi-analytically).
We also compare our results for $\Delta b_4$ at unitarity with Ref.~\cite{YanBlume} (see also Refs.~\cite{Rakshit} and~\cite{Ngampruetikorn2}). We note that, while 
the LO-SCLA is quite rudimentary, the answers it provides are qualitatively correct as a function of $d$ but can be far from the expected numbers (in the sense and scale of Fig.~\ref{Fig:DimensionComparison}). At NLO, on the other hand, the agreement improves considerably for $\Delta b_3$ but again deteriorates for $\Delta b_4$ (when comparing,
in the latter, case, with the only data point available, which is at $d=3$).

\begin{figure}[t]
  \begin{center}
  \includegraphics[scale=0.56]{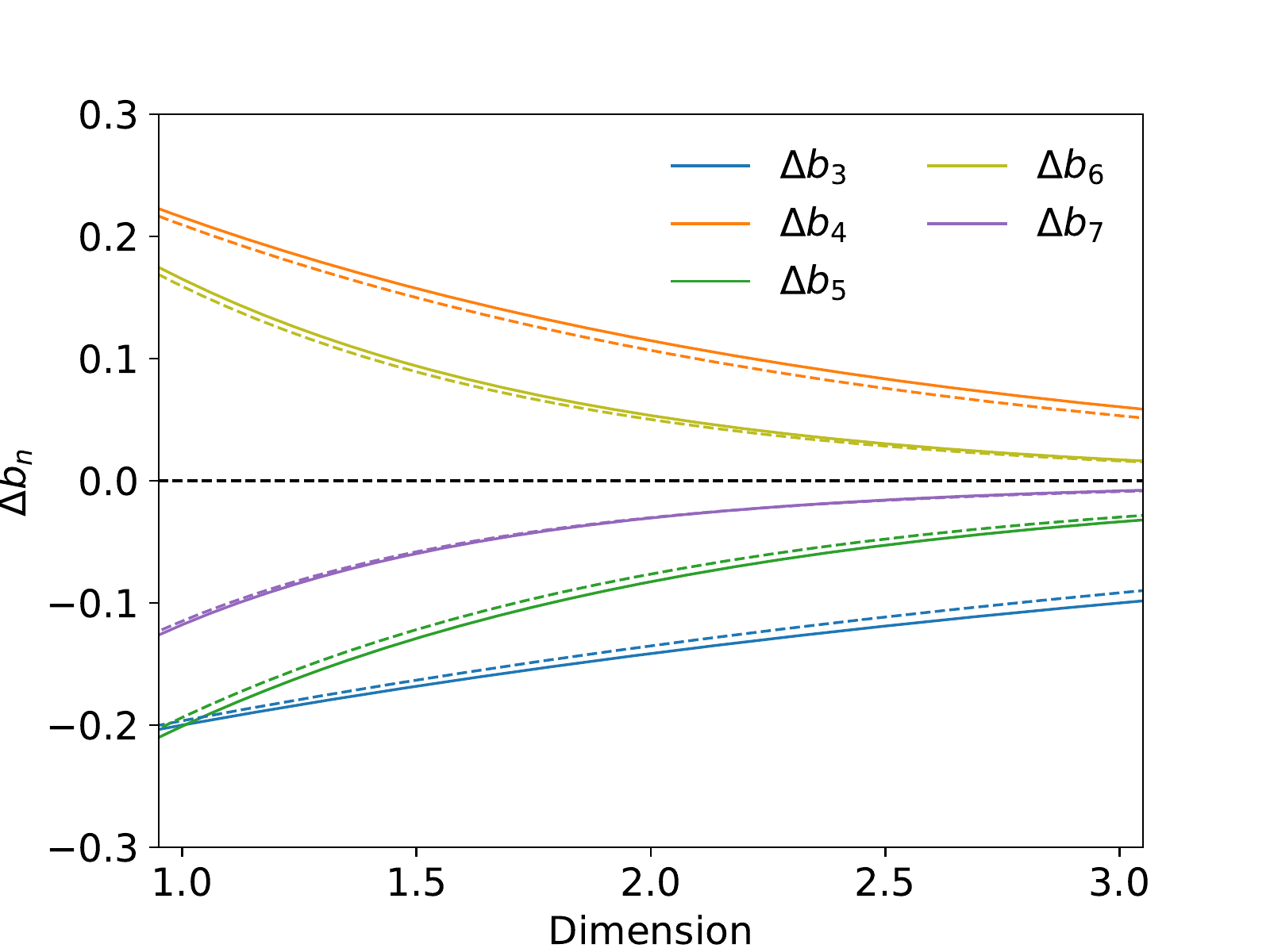}
  \end{center}
  \caption{Interaction-induced change in virial coefficients $\Delta b_3$ -- $\Delta b_7$
  at $\Delta b_2 = 1/(5\sqrt{2})$ (which corresponds to a point on the BCS side of the resonance in 3D where the interaction-induced 
  changes $\Delta b_n$ are of the same order at the noninteracting $b_n$), as a function of the spatial 
  dimension $d$, in the LO- and NLO-SCLA (solid and dashed lines, respectively).
   }
  \label{Fig:DimensionComparisonWeakCoupling}
\end{figure}

While the change in the progression from LO to NLO in Fig.~\ref{Fig:DimensionComparison} is substantial, more weakly coupled regimes than unitarity feature
much improved behavior. As an example, we show a plot of $\Delta b_3, \dots, \Delta b_7$ as functions of the spatial dimension 
$d$ at $\Delta b_2 = 1/(5\sqrt{2})$ in Fig.~\ref{Fig:DimensionComparisonWeakCoupling}, which corresponds to the BCS side of the
3D resonant Fermi gas. In all cases in that figure, the changes when going from LO to NLO are small on the overall scale of the plot,
across all dimensions, whereas the $\Delta b_n$ are of the same order as the noninteracting values $b^{(0)}_n = (-1)^{n+1} n^{-(d+2)/2}$. 
This shows that the SCLA is able to capture the behavior of virial coefficients even in regimes where the interaction effects 
are not small.

\begin{figure}[t]
  \begin{center}
  \includegraphics[scale=0.56]{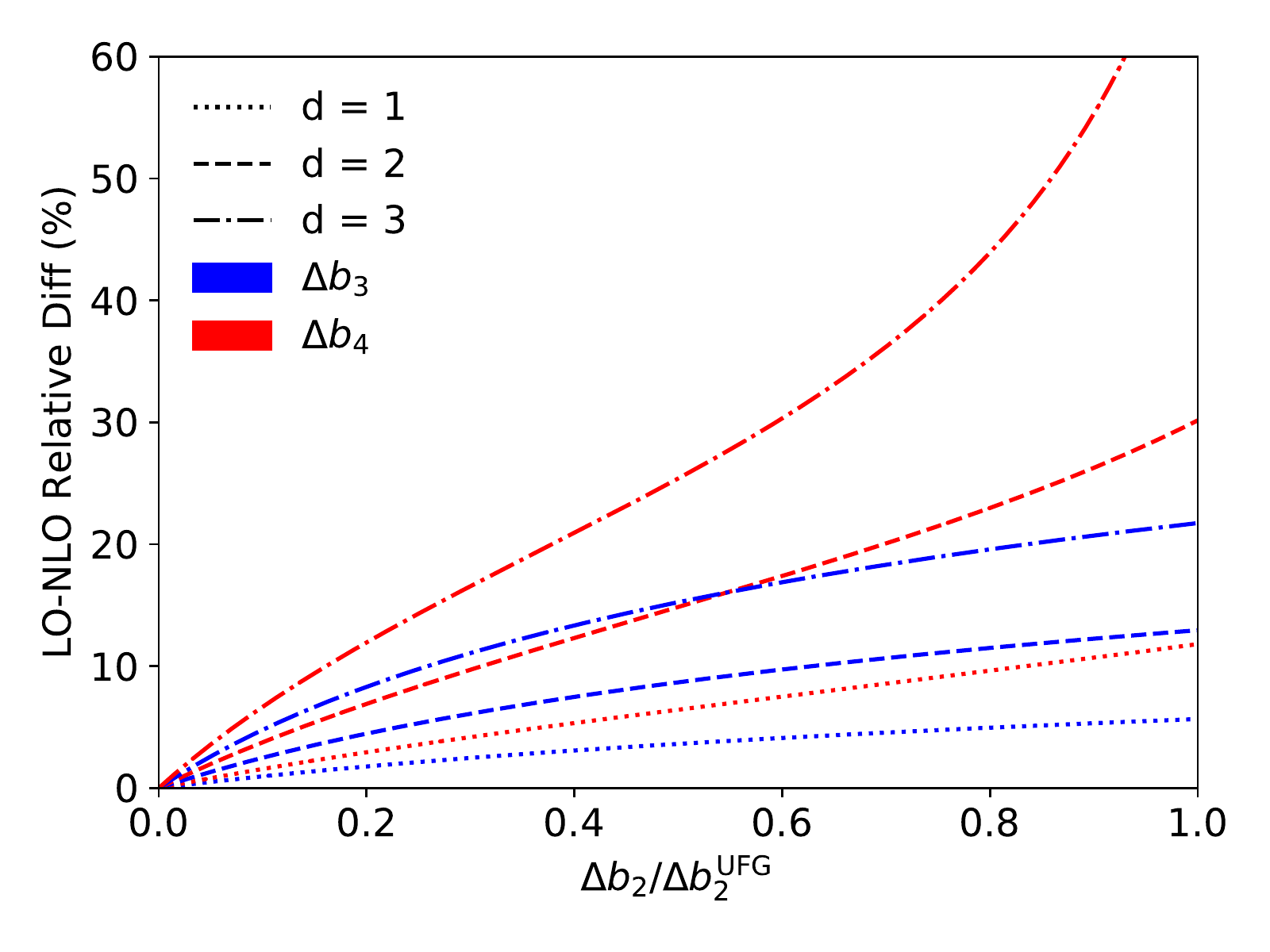}
  \end{center}
  \caption{$|\Delta b_3^\text{NLO}/\Delta b_3^\text{LO} - 1|$ (blue lines) and $|\Delta b_4^\text{NLO}/\Delta b_4^\text{LO} - 1|$ (red lines) as percentages relative to the 
  corresponding LO result, plotted as functions of $\Delta b_2/\Delta b_2^{\mathrm{UFG}}$ for $d=1,2,3$ (in dotted, dashed, and dashed-dotted, respectively). 
   }
  \label{Fig:PercentageComparison}
\end{figure}

To give a more precise sense of the behavior described above, we show in Fig.~\ref{Fig:PercentageComparison} a plot of the LO-NLO change in $\Delta b_3$ and $\Delta b_4$, 
namely $|\Delta b_3^\text{NLO}$-$\Delta b_3^\text{LO}|$ and $|\Delta b_4^\text{NLO}$-$\Delta b_4^\text{LO}|$, as percentages relative to the LO result, 
versus $\Delta b_2 / \Delta b_2^{\mathrm{UFG}}$, where $\Delta b_2^{\mathrm{UFG}} = 1/\sqrt{2}$ corresponds to the unitary limit of the 3D Fermi gas.
Based purely on this LO-NLO analysis,    
we conclude that, in all cases, the convergence properties of the SCLA deteriorate both as a function of the coupling and as a function of the virial order. In other words,
achieving a desired convergence level across a set of $\Delta b_n$ would require higher SCLA orders for higher $n$. This is not unexpected; in fact, it would be surprising
to find the opposite behavior (i.e. improved convergence for higher $n$). Interestingly, lower dimensions display better convergence properties than the higher dimensional 
counterparts across all couplings, which is unexpected given that interaction effects are typically enhanced in low dimensions.

\subsection{Application: pressure and Tan's contact in 2D}

In this section we apply our estimates of the virial coefficients to two simple
thermodynamic observables: the pressure and Tan's contact. For concreteness,
we focus on the 2D attractive Fermi gas, but we emphasize that our
results are explicit analytic functions of the dimension and can therefore be evaluated for 
arbitrary $d$.

\begin{figure}[t]
  \begin{center}
  \includegraphics[scale=0.54]{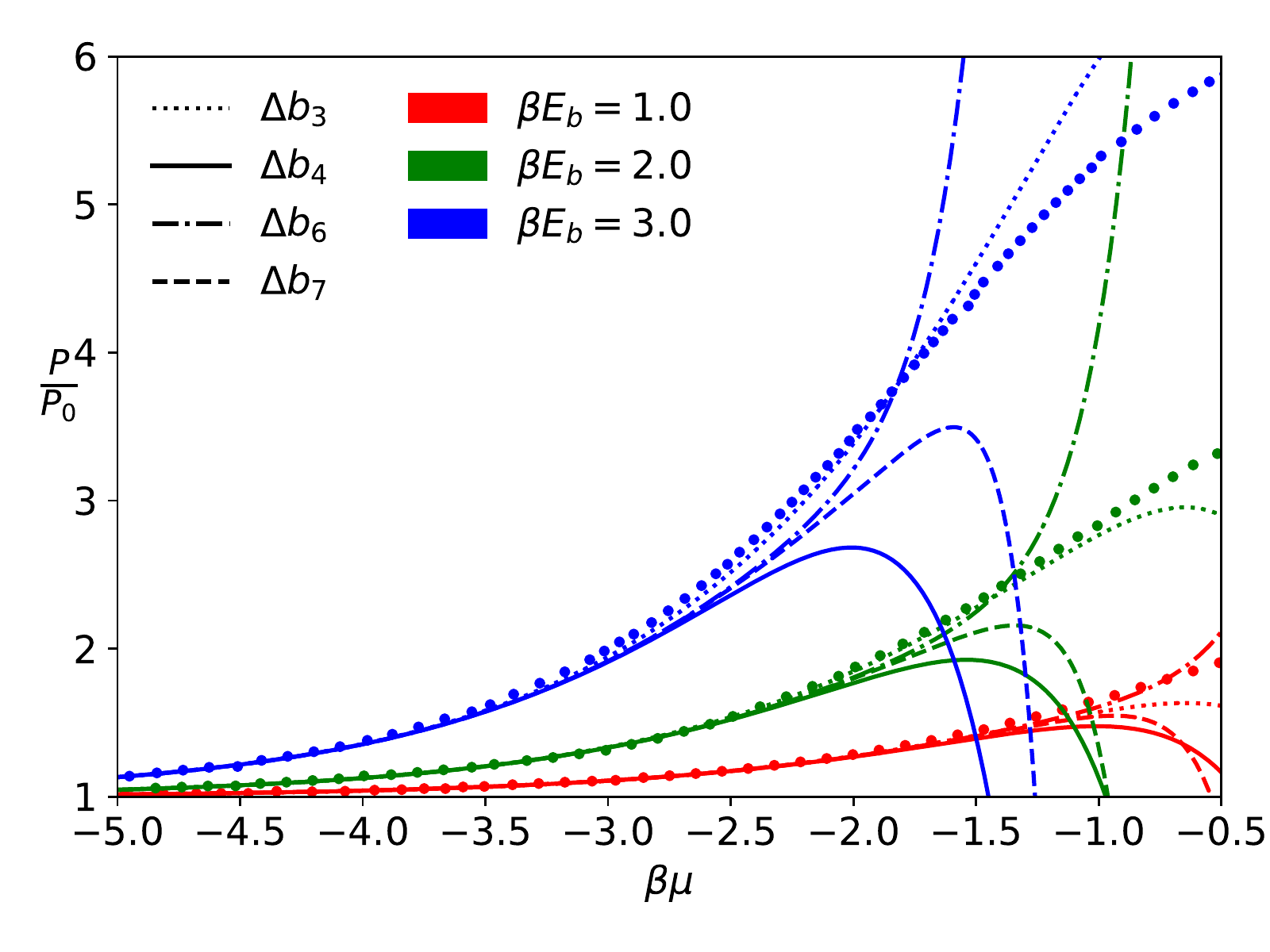}
  \includegraphics[scale=0.54]{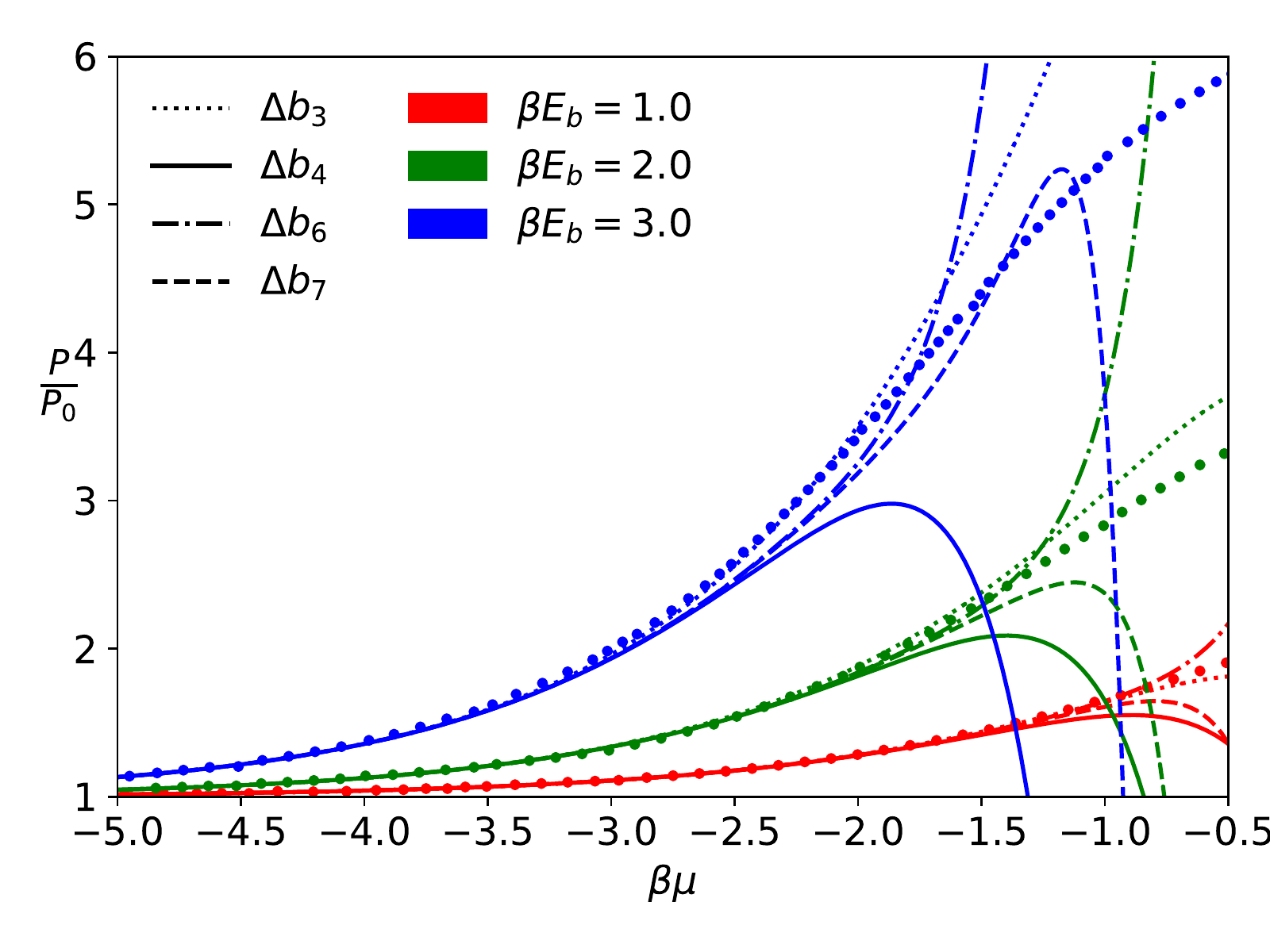}
  \end{center}
  \caption{Pressure $P$ of the 2D attractively interacting Fermi gas, in units of the noninteracting pressure $P_0$,
  both as functions of $\beta \mu$. The lines show our LO-SCLA (top) and NLO-SCLA (bottom) results at various orders in the virial expansion.
  The dots correspond to the quantum Monte Carlo calculations of Ref.~\cite{AndersonDrut}. The colors 
  indicate the value of the coupling: from top to bottom the sets of curves correspond to 
  $\beta \epsilon_B = 3.0$ (blue), $2.0$ (green), and $1.0$ (red).
   }
  \label{Fig:Pressure2D}
\end{figure}

To access the pressure, we combine the calculated virial coefficients according to
\beq
-\beta \Delta \Omega = Q_1 \sum_{m=1}^{\infty} \Delta b_m z^m,
\eeq
where $\Delta \Omega = - \Delta PV$ is the change in the pressure $P$ due to interaction effects. 
From this equation, it is straightforward to determine the density change $\Delta n$ and the compressibility 
change $\Delta \chi$ by differentiation with respect to $z$. In Fig.~\ref{Fig:Pressure2D} we
show the pressure, in units of its noninteracting counterpart $P_0$, as a function of $\beta \mu = \ln z$,
for the 2D Fermi gas with attractive interactions, for the LO- (top) and NLO-SCLA (bottom). The results
at NLO show somewhat improved agreement with previous results at all couplings studied when considering
the full expressions that include $\Delta b_7$.

Within the context of the LO- and NLO-SCLA results for $\Delta b_n$, Fig.~\ref{Fig:Pressure2D} provides an indication of the
range of validity of the virial expansion. In those plots, the large oscillations observed as $\beta \mu$ is increased
show that the radius of convergence $r$ of the virial expansion, as a function of $z$, is notably reduced as
the coupling is increased. For $\beta \epsilon_B = 1$, we have $r\simeq \exp(-1.25)$, whereas for 
$\beta \epsilon_B = 3$, we have $r\simeq \exp(-2.75)$.
Such an effect is expected but, to understand to what extent that reduction is
due to the LO approximation, higher orders in the approximation must be investigated.

\begin{figure}[t]
  \begin{center}
  \includegraphics[scale=0.54]{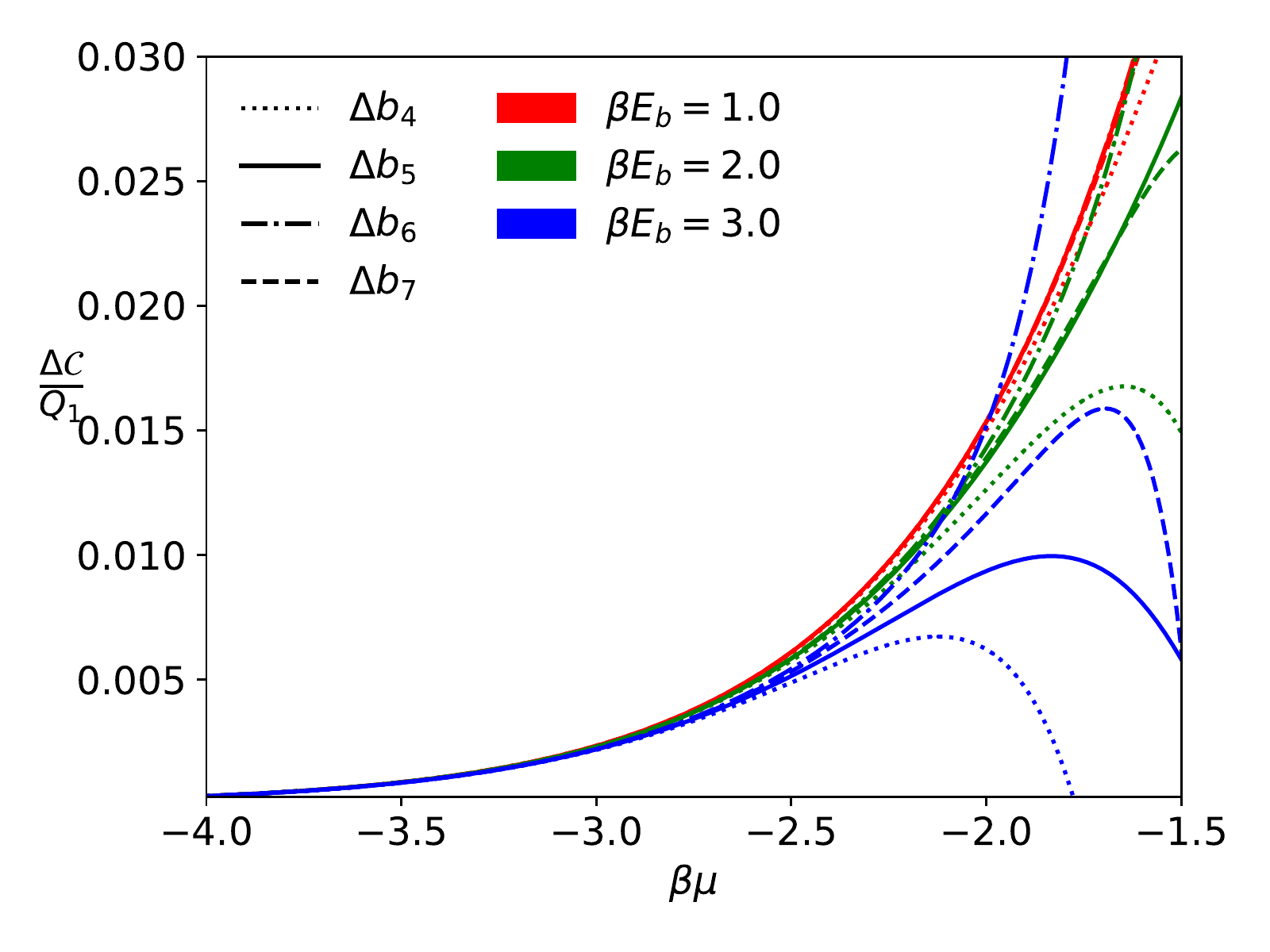}
  \includegraphics[scale=0.54]{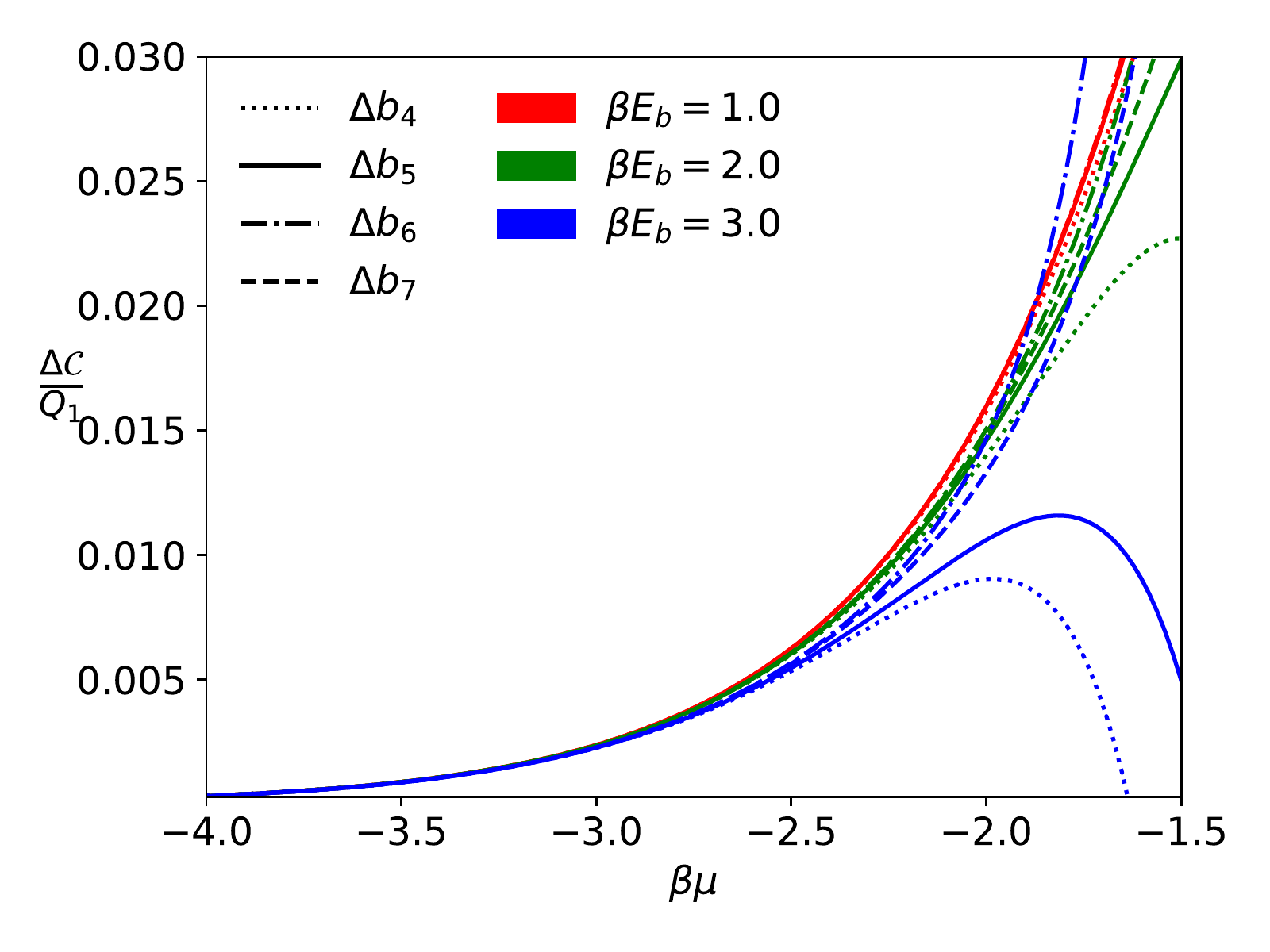}
  \end{center}
  \caption{Many-body contribution to Tan's contact as a function of $\beta \mu$ in the virial expansion, per Eq.~(\ref{Eq:DeltaCQ1})
  The lines show our LO-SCLA (top) and NLO-SCLA (bottom) results at various orders 
  in the virial expansion. The colors indicate the value of the coupling: $\beta \epsilon_B = 3.0$ (blue), $2.0$ 
  (green), and $1.0$ (red).
   }
  \label{Fig:Contact2D}
\end{figure}

To obtain Tan's contact~\cite{Tan}, we differentiate with respect to the coupling $\lambda$ (i.e. we use the so-called adiabatic relation). In this case, since we
use $\Delta b_2$ as our physical dimensionless coupling (or as a proxy to scattering parameters or 
binding energies), we simply differentiate $\Delta \Omega$ with respect to that parameter
and obtain the dimensionless form
\beq
\label{Eq:DeltaCQ1}
\frac{\Delta \mathcal C}{Q_1} = \sum_{m=1}^{\infty} \frac{\partial \Delta b_m}{\partial \Delta b_2} z^m.
\eeq
To connect the above expression to the conventional form of Tan's contact we only need the overall factor
$\partial b_2 / \partial \lambda$, where $\lambda$ is the coupling. Such a factor contains only two-body physics
and can therefore be calculated explicitly using the well-known Beth-Uhlenbeck formula~\cite{BU}.
In Fig.~\ref{Fig:Contact2D} we
show our results for ${\Delta \mathcal C}/{Q_1}$ as a function of $\beta \mu = \ln z$, for the 2D Fermi gas with 
attractive interactions. Although somewhat more difficult to visualize, we glean from this figure that the virial
expansion breaks down at lower values of $\beta \mu$ at stronger couplings.

\section{Summary and Conclusions}

In this work, we calculated the virial coefficients of spin-$1/2$ fermions in the LO-SCLA, 
up to $\Delta b_7$. We have presented analytic results on the lattice and, 
much more succinctly, in the continuum limit, where they feature an explicit analytic dependence 
on the number of spatial dimensions $d$. 

As a renormalization prescription, we fixed the bare constant $C$ by using the fact that $\Delta b_2$ is in many cases 
known analytically through the Beth-Uhlenbeck formula~\cite{BU} (see e.g. Refs.~\cite{EoS1D,virial2D,virial2D2,PhysRevA.89.013614,Daza2D,LeeSchaeferPRC1}). 
That choice allowed us to express our results in powers of $\Delta b_2$ and to perform cross-dimensional comparisons by
varying $d$ at fixed $\Delta b_2$. In turn, that comparison shows that the SCLA behaves qualitatively 
as expected. Notably, the agreement for $\Delta b_3$ improves across dimensions when going from LO to NLO,
but it appears to deteriorate for $\Delta b_4$ at unitarity. To better understand that feature, we showed results at
weaker couplings, which show much better convergence (i.e. much smaller changes) when going from LO to NLO
for all the $\Delta b_n$ studied. These results are encouraging towards exploring higher orders in the SCLA, which 
will be carried out elsewhere~\cite{HouDrut}.

The fact that we have used the continuum limit of momentum sums in our final answers, together with the continuum $\Delta b_2$ result to fix the coupling, 
does not eliminate all the lattice artifacts. Indeed, implicit in our derivations is the spatial lattice spacing $\ell$, with which the coupling runs and
which induces finite-range effects. To avoid such effects, future studies should use improved actions (see e.g.~\cite{DrutNicholson, Drut}).

Finally, it should be stressed that the applicability of the SCLA goes beyond the approximation of virial coefficients. In the form implemented here,
it can be used to access the thermodynamics of finite systems. For those, the SCLA simply represents the non-perturbative analytic evaluation of a 
finite-temperature lattice calculation which would not be possible due to the sign problem, and which is carried out in a coarse temporal lattice.


\acknowledgments
This material is based upon work supported by the National Science Foundation under Grant No.
PHY{1452635} (Computational Physics Program).

\appendix

\section{Matrices}

In this section we provide the detailed form of the matrices that appear in the 
evaluation of the functions in Tables~\ref{Tab:Db2Db7C1SCLALattice},~\ref{Tab:Db2Db7C2SCLALattice}, 
and~\ref{Tab:Db2Db7C3SCLALattice}. 
To that end, it is useful to define the following notation:
\beq
    M_0 = 
    \begin{pmatrix}
      0 & 1 & -1 \\
      1 & 0 & -1 \\
      -1 & -1 & 0 \\
    \end{pmatrix},
    \ \ \ \ \ 
    \text{diag}(a,b,c) = 
    \begin{pmatrix}
      a & 0 & 0 \\
      0 & b & 0 \\
      0 & 0 & c \\
    \end{pmatrix}.
\eeq
Using the above, we have
\bea
M_4 &=& \text{diag}(2,2,2) + M_0 \\ 
M_5 &=& \text{diag}(3,2,2) + M_0 \\
M_{6a} &=& \text{diag}(4,4,4) + 3 M_0 \\
M_{6b} &=& \text{diag}(3,2,3) + M_0 \\
M_{6c} &=& \text{diag}(4,3,3) + 2M_0 \\
M^{(2)}_{7a} &=& \text{diag}(5,4,4) + 3M_0 \\
M^{(2)}_{7b} &=& \text{diag}(5,2,2) + M_0 \\
M^{(2)}_{7c} &=& \text{diag}(4,4,3) + 2M_0 \\
M^{(3)}_{7a} &=& \text{diag}(2,3,2) + M_0 \\
M^{(3)}_{7b} &=& \text{diag}(3,3,2) + M_0 \\
M^{(3)}_{7c} &=& \text{diag}(4,4,4) + 3M_0
\eea

\bea  
\label{eq:2}
    N_1 = 
    \begin{pmatrix}
      3 & 2 & -1 & -1 \\
      2 & 3 & -1 & -1 \\
      -1 & -1 & 2 & 0 \\
      -1 & -1 & 0 & 2 \\
    \end{pmatrix}
\eea

\bea  
    N_2 = 
    \begin{pmatrix}
      4 & 2 & -1 & -1 \\
      2 & 3 & -1 & -1 \\
      -1 & -1 & 2 & 0 \\
      -1 & -1 & 0 & 2 \\
    \end{pmatrix}.
\eea

\section{High-order virial expansion formulas}

For completeness, we provide here some of the formulas which we omitted in the main text for the sake of brevity and clarity.
These are model independent, except as noted below. The complete expressions for $b_5$, $b_6$, and $b_7$ in terms of the 
corresponding canonical partition functions and prior virial coefficients can be written as
\bea
Q_1 b_5 &=& Q_5 - (b_4  + b_2 b_3 ) Q_1^2 - \left (b_2^2  + b_3 \right )\frac{Q_1^3}{2} \nonumber \\
&&- b_2 \frac{Q_1^4}{3!}  - \frac{Q_1^5}{5!},  \\
Q_1 b_6 &=& Q_6 - \left ( b_5 + \frac{b_3^2}{2}  + b_2 b_4 \right)Q_1^2 \nonumber \\
&& - \left( \frac{b_2^3}{6} + \frac{b_4}{2}+  b_3 b_2 \right ) Q_1^3 \nonumber \\
&& - \left (\frac{b_2^2}{2} +  \frac{b_3}{3} \right) \frac{Q_1^4}{2!}  - b_2\frac{Q_1^5}{4!}    - \frac{Q_1^6}{6!}, \\
Q_1 b_7 &=& Q_7 - (b_6 + b_2 b_5 + b_3 b_4) Q_1^2 \nonumber \\
&& - \left (\frac{b_3^2}{2} + \frac{b_5}{2}+ b_2  b_4 + \frac{b_3 b_2^2}{2}\right) Q_1^3 \nonumber \\
&& - \left(\frac{b_2^3}{6} + \frac{b_4}{6} + \frac{b_3 b_2}{2}\right) Q_1^4 \nonumber \\
&& - \left(b_2^2 + \frac{b_3}{2}\right)\frac{Q_1^5}{12} - b_2\frac{Q_1^6}{5!}   - \frac{Q_1^7}{7!},
\eea
whereas the change in the above due to interactions (assuming here two-body interactions) are given by
\begin{widetext}
\bea
Q_1 \Delta b_5 &=&\Delta  Q_5 - \Delta (b_4  + b_2 b_3 ) Q_1^2 - \frac{1}{2}\Delta \left (b_2^2  + b_3 \right )Q_1^3 - \frac{\Delta b_2}{3!} Q_1^4,  \\
Q_1 \Delta b_6 &=& \Delta Q_6 - \Delta \left ( b_5 + \frac{b_3^2}{2}  + b_2 b_4 \right)Q_1^2 - \Delta \left( \frac{b_2^3}{6} + \frac{b_4}{2}+  b_3 b_2 \right ) Q_1^3 
 - \Delta \left (\frac{b_2^2}{4} +  \frac{b_3}{6} \right) Q_1^4  - \frac{\Delta b_2}{4!}  Q_1^5 , \\
Q_1 \Delta b_7 &=& \Delta Q_7 - \Delta (b_6 + b_2 b_5 + b_3 b_4) Q_1^2 - \Delta \left (\frac{b_3^2}{2} + \frac{b_5}{2}+ b_2  b_4 + \frac{b_3 b_2^2}{2}\right) Q_1^3 
 - \Delta \left(\frac{b_2^3}{6} + \frac{b_4}{6} + \frac{b_3 b_2}{2}\right) Q_1^4 \\ 
 && - \Delta \left(\frac{b_2^2}{12} + \frac{b_3}{24}\right)Q_1^5 - \frac{\Delta b_2}{5!} Q_1^6.
\eea
\end{widetext}
To use these, it is useful to have the following:
\bea
\Delta (b_n)^2 &=& \Delta(b_n^2) + 2b_n^{(0)} \Delta b_n , \\
\Delta (b_n)^3 &=& \Delta(b_n^3) + 3 b_n^{(0)} \Delta(b_n^2) + 3 (b_n^{(0)})^2 \Delta b_n,
\eea
\bea
\Delta (b_n b_m) &=& \Delta b_n  \Delta b_m + b_n^{(0)} \Delta b_m + b_m^{(0)} \Delta b_n, \\
\Delta (b_n b_m^2) &=& \Delta b_n (\Delta b_m)^2 + b_n^{(0)} (\Delta b_m)^2 + 2 b_m^{(0)} \Delta b_n \Delta b_m \nonumber \\
&&\!\!\!\!\!\!\!\!  + 2 b_n^{(0)} b_m^{(0)} \Delta b_m + (b_m^{(0)})^2 \Delta b_n.
\eea


\end{document}